 \documentclass[prl,aps,twocolumn,showpacs]{revtex4}
\usepackage[dvips]{graphicx}
\usepackage{dcolumn}
\newcommand{\be}{\begin{equation}}
\newcommand{\ee}{\end{equation}}
\newcommand{\bea}{\begin{eqnarray}}
\newcommand{\eea}{\end{eqnarray}}
\newcommand{\nn}{ \nonumber}

\begin{document}
\topmargin=-15mm

\title{ Nature of the intensification of a cyclotron resonance in potassium in a normal magnetic field}

\author{N. A. Zimbovskaya and V. I. Okulov }

\affiliation
{Institute of Metal Physics, Ural Science Center, Academy of Sciences of the USSR  }

\begin{abstract}
 The cyclotron-resonance peak which has been observed [G. A. Baraff {\it et al.}, Phys. Rev. Lett. {\bf 22}, 590 (1969) ]  in potassium  in a magnetic field directed perpendicular to the surface  may be due to an effect of zero-curvature points of the Fermi surface on the cyclotron orbit of effective electrons.
 \end{abstract}

\maketitle

Recent experimental data \cite{2,3} have altered the picture of the Fermi suface of potassium as a closed, nearly spherical surface. That picture had been drawn on the basis of an analysis of de Haas-van Alphen oscillations \cite{4}. Coulter and Datars \cite{2} have observed open orbits for several directions of the magnetic field. Jensen and Plummer \cite{3} have discussed photoemission data which imply the existence of small energy gaps in potassium, as in sodium. According to the explanation offered for these results \cite{2,5,6}, there are charge density waves in the ground state of the conduction electrons in potassium. These waves lead to discontinuities of spherical Fermi surface at several planes.

The observed distortions of the Fermi sphere should be accompanied by the occurrence of transitions from a positive curvature to a negative curvature, i.e., by the presence of points or lines of zero curvature. Such points or lines will lead to characteristic effects in the dispersion and absorption of short-wave sound \cite{7,8}. They should also affect the frequency dependence of the surface impedance and the cyclotron resonance in a normal field \cite{9}. It is thus necessary to reexamine the explanation offered for the experimental results of Ref. 1, where a cyclotron-resonance peak was observed in a potassium plate in a normal field. The theory for a resonance of this sort, which is based on the assumption that the Fermi surface is spherical \cite{10}, explains neither the comparatively large amplitude of the peak nor its shape. It can now be suggested that an intensification of the resonance occurs because zero-curvature points fall on the cyclotron orbit of the effective electrons, as predicted in Ref. \cite{9}. In offering an explanation for the intensification of the resonance, Lacueva and Overhauser \cite{11} assumed that this resonance stems from a small part of the Fermi surface, with the shape of a right cylinder, which has been split off by a gap. However, the assumption that there exists a finite, strictly cylindrical part of the Fermi surface, with the same cyclotron frequency as on a sphere, is not justified. Furthermore, a resonance would occur in that model only if the magnetic field were oriented along the axis of the cylinder. This point makes it difficult to explain the experimental data obtained with a polycrystalline sample. There is a stronger case for regarding the amplification of the resonance as resulting from a local vanishing of the curvature along the basic part of the Fermi surface. Since there are many gaps, the points of zero curvature are distributed quite widely and fall on the cyclotron orbit of the effective electrons over considerable intervals of the field orientation. There will then be a marked increase in the effective mass corresponding to motion away from the boundary and thus in the time spent by electrons in the skin layer; the effect will be to intensify the resonance. The deviations from a spherical shape of the Fermi surface are, on the whole, small and do not substantially change the cyclotron frequency. The observed cyclotron-resonance peak can be described on the basis of arguments of this sort.

In calculating the conductivity $ \sigma $ for a resonant circular polarization of an alternating field of frequency $ \omega $ under conditions corresponding to the anomalous skin effect, the deviations of the Fermi surface from a spherical shape are important only for the component of the electron velocity which runs normal to the boundary, $v_z. $ We can accordingly write the conductivity in a normal field as follows \cite{9}:
    \be 
 \sigma = \frac{ie^2}{8 \pi^3\hbar^3 q} \int d \psi \int d p_z \frac{m v_\perp^2 (p_z)}{w - v_z(p_z,\psi)} \approx
\frac{e^2}{4 \pi \hbar^3 q} p_F^2 (1 + s),
   \ee
  where $ m $ and $ v_\perp (p_z) $ are the cyclotron mass and transverse velocity component on a spherical Fermi surface; $ w = (\omega - \Omega + i\nu)/q; \ \Omega $ is the cyclotron frequency; $ \nu $ is the collision frequency; and $ q $ is a wave vector. The integration in (1) is carried out over the momentum projection $ p_z $ and over the angle $ \psi = \Omega t \ (t $ is the time of motion along the orbit). The asymptotic behavior of the integral is calculated in the limit $ w \to 0. $ For a spherical Fermi surface, $ v_z $ is independent of $ \psi , $ and we have $ s = 0. $ The quantity $ s $ describes the contribution of a point of zero curvature if it corresponds to effective electrons with $ v_z (p_z,\psi) = 0. $ The minimum of the function $ v_z (p_z, \psi) $ under the conditions $ p_z = p_{z0} $ and $ \psi = \psi_0 $ would be one possibility for a point of this type. Near it, for small values of $ p_z - p_{z0}, \ \psi - \psi_0, $ we can write
   \bea 
   v_z(p_z,\psi) &=& a(p_z - p_{z0})^2 + 2b(p_z - p_{z0})(\psi - \psi_0) \nn\\ & +& c(\psi - \psi_0)^2; \qquad \ \ \
 a>0; \nn\\&& ac - b^2 >0.
   \eea
 The corresponding value of $ s $ is 
  \be 
 s = \eta (\pi - i\ln w_0/w),
  \ee
\begin{figure}[t]
\begin{center}
\includegraphics[width=5.5cm,height=7cm]{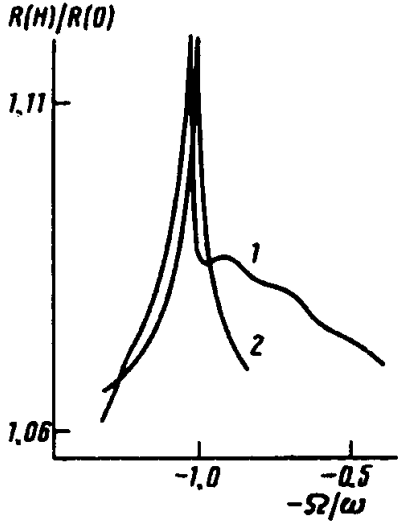}
\caption{Curve 1 is relative magnitude of the real part of the impedance versus the magnetic field (the ratio $ \Omega/\omega = H/H_r) $ near the resonance found in the experiments by Baraff {\it et al} \cite{1}; curve 2 is plot of expression (4) with $ \gamma^2 = 2 \times 10^{-4} $ and $ \eta = 0.03. $ The origin of the scale for the measured values of $ R(H)/R(0), $ which was not determined experimetally, has been chosen in a way different from that in the figure in Ref. \cite{1} -- in such a way that the resonant values are identical for curves 1 and 2. The cyclotron frequency $ \Omega $ has been determined for a resonance in a field parallel to the boundary. A possible explanation for the slight difference between the positions of the peaks on curves 1 and 2 is that the effective electrons have a slightly shifted resonant frequency in a normal field.
}
\label{rateI}
\end{center}
\end{figure}
 where the parameters $ \eta $ and $ w_0 $ characterize the properties of the Fermi surface near the point of zero curvature. In particular, $ \eta $ is, in order of magnitude, the relative size of the region in which dependence (2) holds. Since the distortions of the Fermi sphere are small, we have $ \eta \ll 1; $ in calculating the surface impedance from (1)--(3) we can therefore expand it in powers of $ \eta. $ The resonant increment in which we are interested turns out to be small, in accordance with the experimental results of Ref. \cite{1}. We write, in the linear approximation in $ \eta, $ the result calculated for the ratio of the real part of the impedance in a magnetic field, $R(H)$ to that without the field, $ R(0) $ (this ratio was measured in Ref. \cite{1}):
   \be 
  \frac{R(H)}{R(0)} = 1 - \frac{\eta}{3} \Big[ \sqrt3 \ln \sqrt {\Delta^2 + \gamma^2} + \mbox{sign} \Delta \arctan \frac{\gamma}{\sqrt{\Delta^2 + \gamma^2}} \Big],
  \ee
  where $ \Delta = 1 - H/H_r, \ H_r $ is the resonant value of the field, and $ \gamma = \nu/\omega \ll 1. $ Expression (4) describes a positive, asymmetric, resonance peak with an abrupt low-field cutoff and a slow decay with distance from the point of the resonance. The peak observed in Ref. \cite{1} has specifically this qualitative shape. A particularly important point is that the height of the peak, which is not explained in the model of a spherical Fermi surface, can be reconciled with (4) at a plausible value of $ \eta. $ The scale of the variations and the height of the peak near the resonance $(|\Delta| \leq 0.1)$ are close to those observed at values of $ \eta $ and $ \gamma $ on the order of $10^{-1}\div 10^{-2} $ (Fig. 1).

A resonance may also be caused by the appearance of zero-curvature points of other types. For example, a narrower peak arises in the case in which we have  $ ac- b^2 \to 0 $ in Eq. (2). Simulating the function $ v_z(p_z, \psi )$ by the functional dependence $ v_z (p_z, \psi) = a[p_z - p_{z0} (\psi)]^2$ over a small but nonzero interval of $ \psi $ for this case, we find $ s = \eta \sqrt{w_0/w}. $ The $\Delta$-dependence factor in the term proportional to $ - \eta $ in Eq. (4) takes the form $ \mbox{Re} \big \{1/\sqrt{\Delta + i \gamma} \big \} + (2 + \sqrt 3) \mbox{Im} \big \{1/\sqrt{\Delta + i \gamma}\big \}. $

Expression (4) was derived without allowance for the Fermi-liquid interaction or the surface scattering of electrons. These factors may influence the position and intensity of the resonance peak, so this peak may shift slightly and change in shape when these factors are taken into consideration. However, the basic result will remain in force. A cyclotron resonance is amplified in a normal field because of the existence of zero-curvature points.
\vspace{2mm}

{\it Acknowledgments:}
NAZ thanks G. M. Zimbovsky for help with the manuscript.

\end{document}